\documentclass[11pt,twoside]{article}
\usepackage{asp2010}

\resetcounters

\begin{document}

\title{Modelling the spectra of (BAL)QSOs}
\author{Nick~Higginbottom,$^1$ Knox~S.~Long,$^2$ Christian~Knigge,$^1$ and Stuart~A.~Sim$^3$
\affil{$^1$School of Physics and Astronomy, University of Southampton, University Road, Southampton SO17 1BJ, England}
\affil{$^2$Space Telescope Science Institute, 3700 San Martin Drive, Baltimore MD 21218, USA}
\affil{$^3$Research School of Astronomy and Astrophysics, Mount Stromlo Observatory, Cotter Road, Weston Creek, ACT 2611, Australia}}

\begin{abstract}
We have embarked upon a project to model the UV spectra of BALQSOs using a Monte Carlo radiative transfer code previously validated through modelling of the winds of cataclysmic variable stars \citep[e.g.][]{2010ApJ...719.1932N}. We intend to use the simulations to investigate the plausibility of geometric unification \citep[e.g.][]{2000ApJ...545...63E} of the different classes of QSO. Here we introduce the code, and present some initial results. These demonstrate that for reasonable geometries and mass loss rates we are able to produce synthetic spectra which reproduce the important features of observed BALQSO spectra. 
\end{abstract}

\section{Motivation}
Broad absorption line quasars (BALQSOs) are a class of QSO exhibiting broad, blue shifted absorption lines of highly ionized species such as N~{\sc v}, C~{\sc iv}, O~{\sc vi} and Si~{\sc iv}. These objects make up perhaps 17\% of the population of QSOs \citep[e.g.][]{2008MNRAS.386.1426K, 2003AJ....125.1784H}. Their unusual UV spectra are believed to be the result of fast (0.01-0.2c) winds originating from the accretion disk \citep[e.g.][]{1992ApJ...401..529K}, although the geometry and kinematics of these winds are not well understood, and even the driving mechanism is unclear.  
It is a source of discussion as to whether this observation demonstrates that all QSOs spend 17\% of their lifetime as a BALQSO (`evolutionary unification'; e.g. \citealt{2000ApJ...538...72B}), or whether all QSOs have these observational signatures when observed from 17\% of possible directions (`geometrical unification';  e.g. \citealt{2000ApJ...545...63E}). Our aim in developing a code to simulate UV spectra for quasar disk winds is to permit quantitative evaluation of prospective geometries for geometrical unification.

\section{Python - A Radiative Transfer code}
Python is a Monte Carlo type radiative transfer code, first described in \cite{2002ApJ...579..725L}, which has been used previously to simulate the spectra of cataclysmic variables \citep[e.g.][]{2010ApJ...719.1932N} and massive young stellar objects \citep{2005MNRAS.363..615S}. Cataclysmic variables are thought to drive winds with similar geometries, kinematics and ionization conditions to those in AGN and are observed to produce similar wind-formed spectral lines and line profile shapes. It is therefore natural to attempt to apply Python to the modelling of BALQSO spectra and hence, if geometrical unification is possible, QSO spectra in general.

Python is designed to be easy to adapt to different kinematic descriptions of winds. A number of azimuthally symmetric wind models are currently implemented in the code, including the wind prescriptions of \cite{1993ApJ...409..372S}, \cite{1995MNRAS.273..225K}, and one simulating the wind envisioned by \cite {2000ApJ...545...63E}. 

\begin{figure}
\plotone{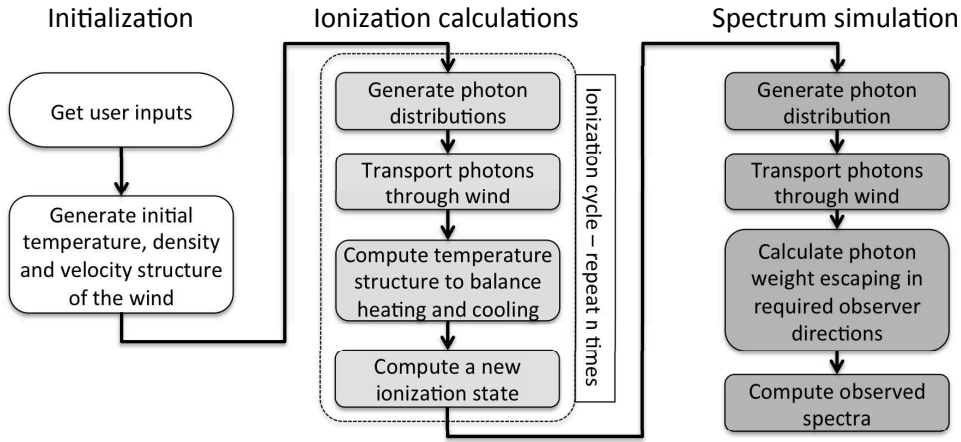}
\caption{Schematic of Python execution cycle}
\label{pyflow}
\end{figure}

Figure \ref{pyflow} shows schematically the execution stages of the code.
Once the wind model is defined, and the computation grid populated by wind cells of given density and velocity, the code proceeds by generating a population of photons arising from all radiating parts of the simulated system. These can include:
\begin{itemize}
\item{the disk - a thin disk approximation is used here, implemented as a series of annuli radiating as black bodies (in principle any spectral energy distribution can be used);}
\item{the wind - radiation is produced via line and continuum processes at the current electron temperature;}
\item{a spherical `corona' at the centre of the geometry - producing photons defined by a power law extending into the x-ray regime. In the current implementation this region is contained within the innermost extent of the disk.}
\end{itemize}
The initial ionization state of the wind is calculated using the Saha equation and an initial estimate of the electron temperature. Photons are then tracked through the wind. These photons are used to calculate the local radiation field for each cell, and this is then used to compute new ionization state. This process is repeated until the temperature and ionization state have converged throughout the entire outflow. 

The ionization calculations are optimised for speed to allow even complex wind models and radiation sources to be considered. Key to this is a parametrisation of the photon field in any given cell. This parametrisation is then used to solve for the ionization state via a modified version of the Saha equation. All of the previous work using Python has been based on a dilute blackbody approximation \citep{mazzali_lucy}. However, it is unclear whether this method is appropriate for an AGN exhibiting a strong power-law component. Therefore we have recently incorporated a broken power-law approximation to the local radiation field \citep{2008MNRAS.388..611S} into Python. We are currently experimenting with this to determine whether it will improve the estimate of the ionization state in AGN winds. 
\section{Wind Geometry}
\begin{figure}[h]
\centering
\plotone{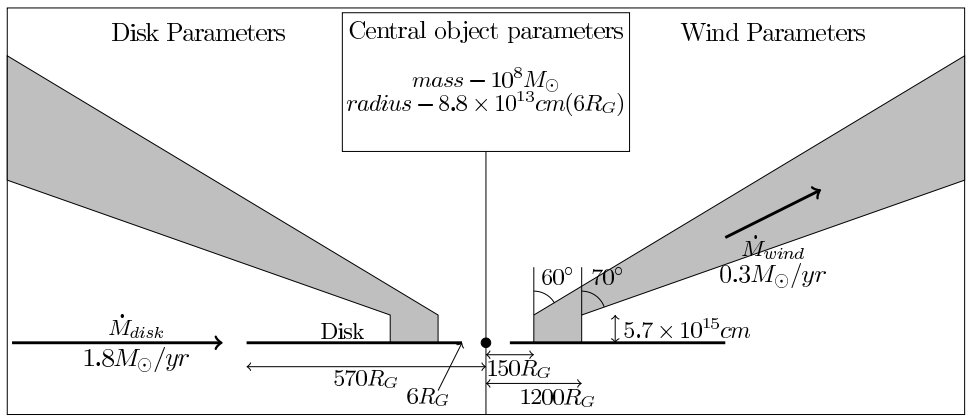}
\caption{Cartoon of the wind geometry for one possible Elvis-style wind}
\label{wgeom}
\end{figure}

Here we present simulated AGN spectra based on the geometry suggested by \cite{2000ApJ...545...63E}. A schematic of the wind parameters is given in Figure \ref{wgeom} together with the values of the most important model parameters. Parameters controlling the radial dependance of the mass loading of the wind and the wind acceleration are omitted for brevity.
\section{Example results}
\begin{figure}[!ht]
\centering
\plottwo{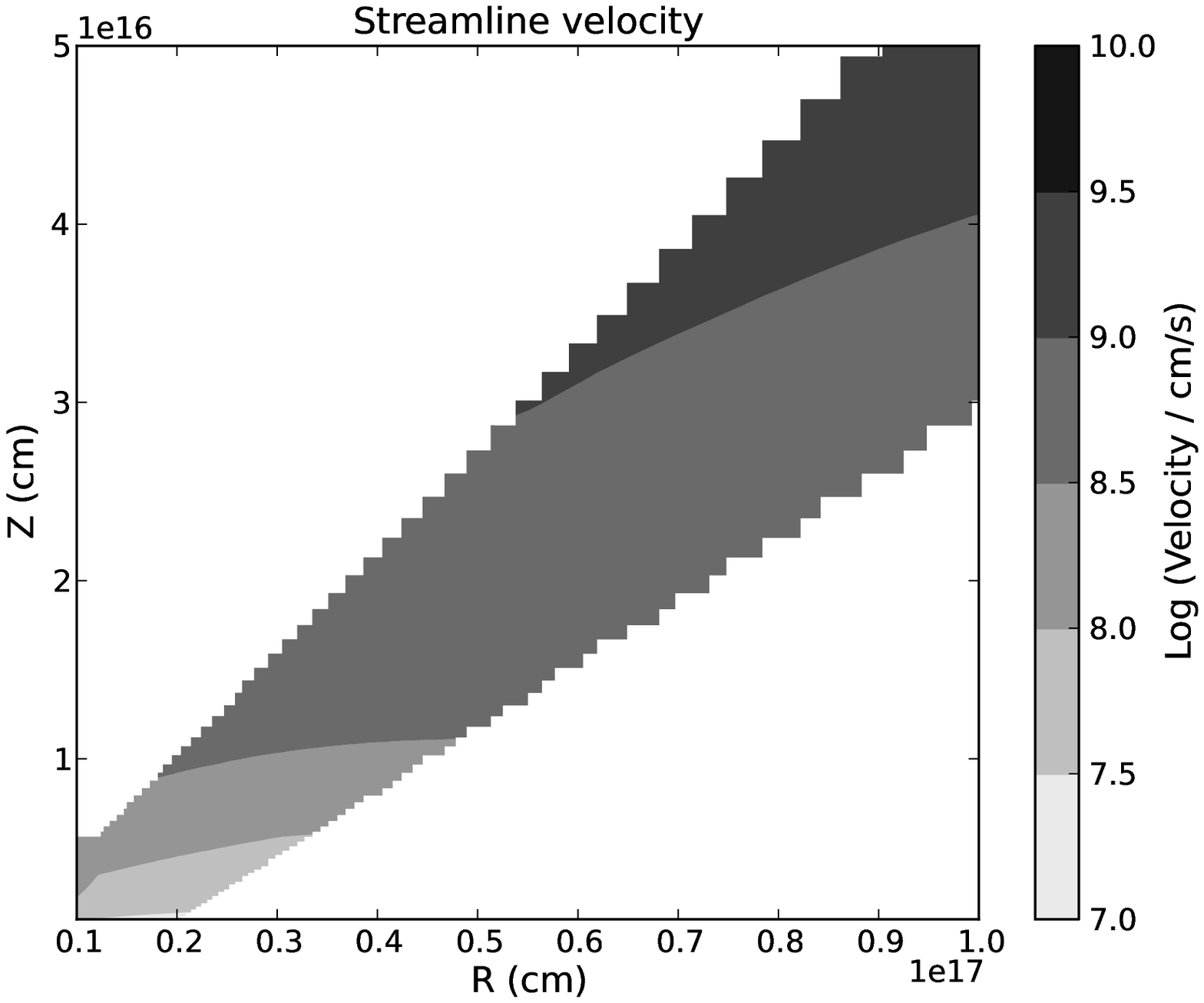}{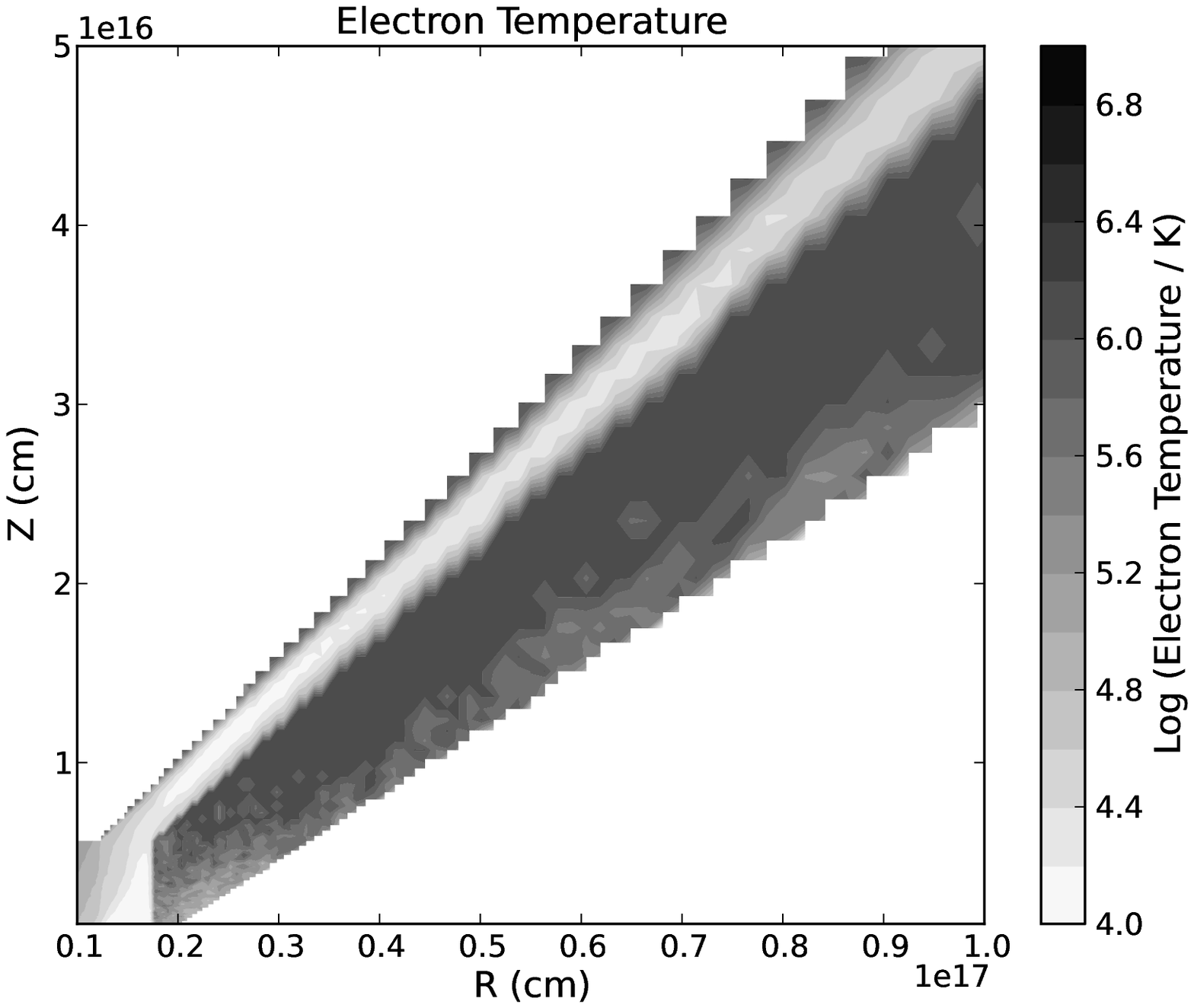}
\plottwo{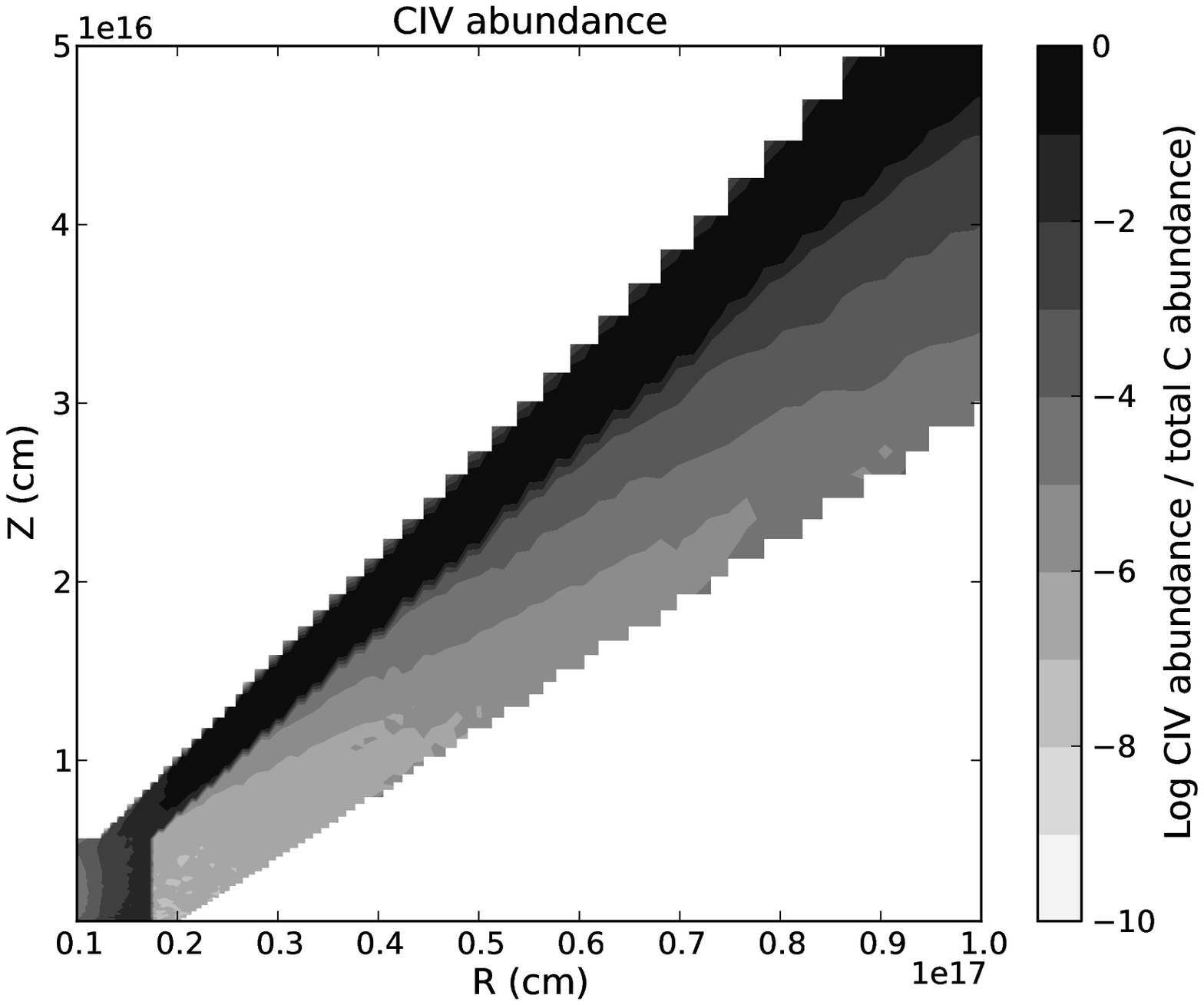}{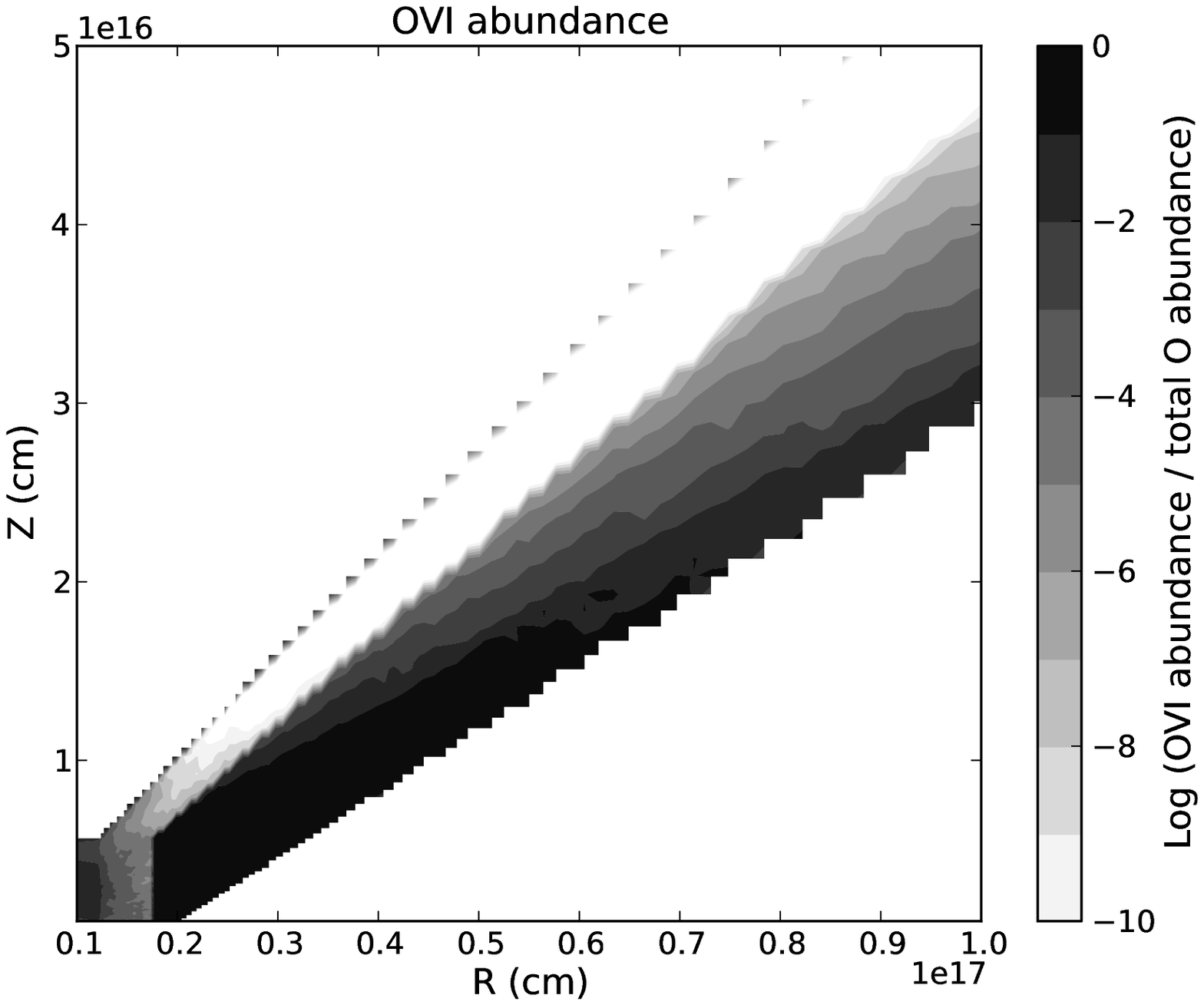}
\caption{Wind velocity, calculated electron temperatures, C~{\sc iv} and  O~{\sc vi} abundances for the model shown in Figure \ref{wgeom}}
\label{tevel}
\end{figure}

In Figure \ref{tevel} we present some results from the geometry described above, illuminated with photons generated from the disk only. The top left plot shows the assigned wind velocity, and the the electron temperature calculated for the wind in the top right plot.

The bottom left plot shows the calculated fractional abundance of C~{\sc iv} 1550\AA~(the most ubiquitous and often strongest BAL feature) and the fracional abundance of O~{\sc vi} 1240\AA~ is given in the bottom right plot. These plots show that, in this model, O~{\sc vi} is found in the lower part of the wind, where the temperature is high, whilst we find C~{\sc iv} in the upper, cooler part of the wind. Also note that the velocity of the wind is higher in the upper part of the wind where the C~{\sc iv} is found.

\begin{figure}[!ht]
\centering
\plottwo{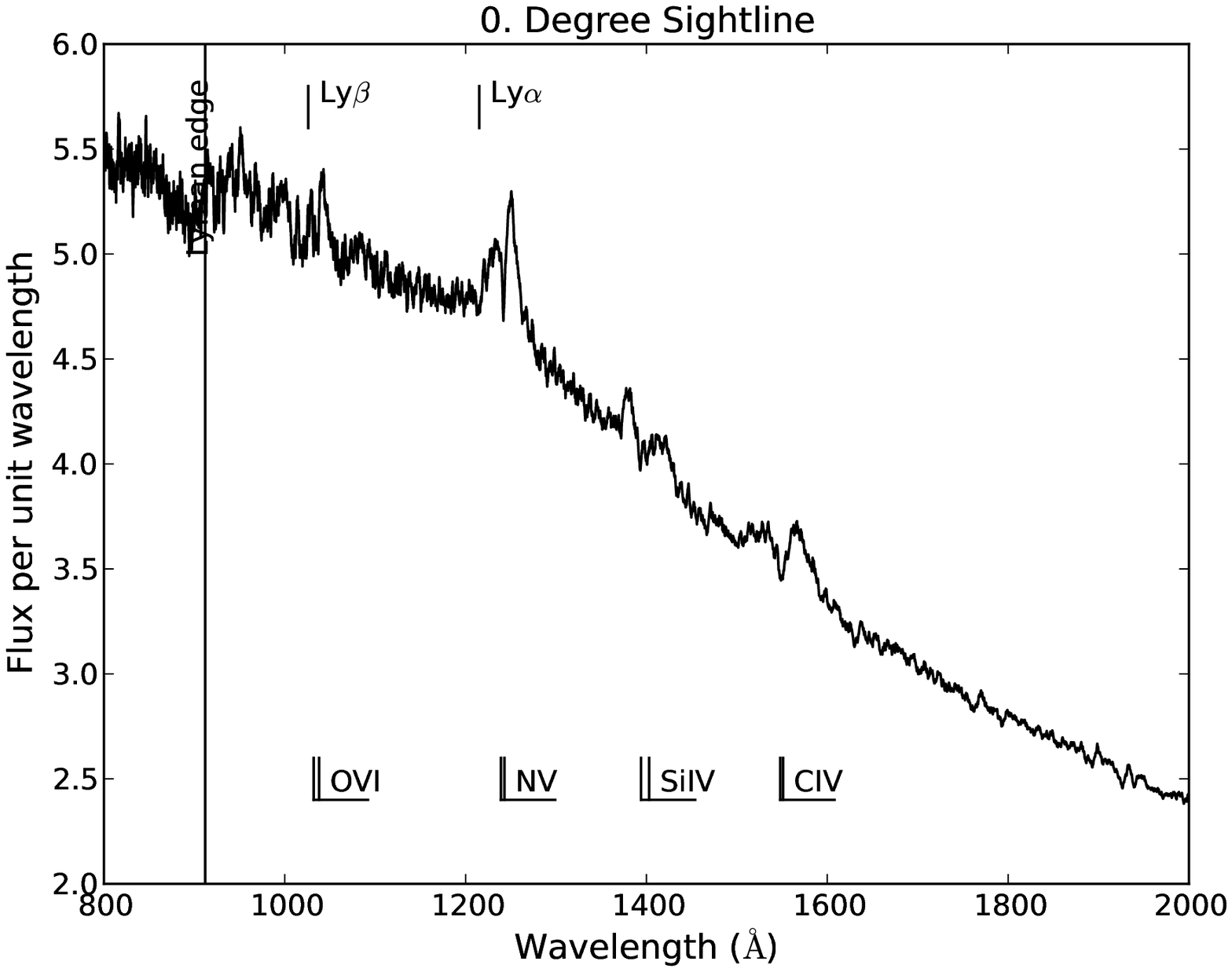}{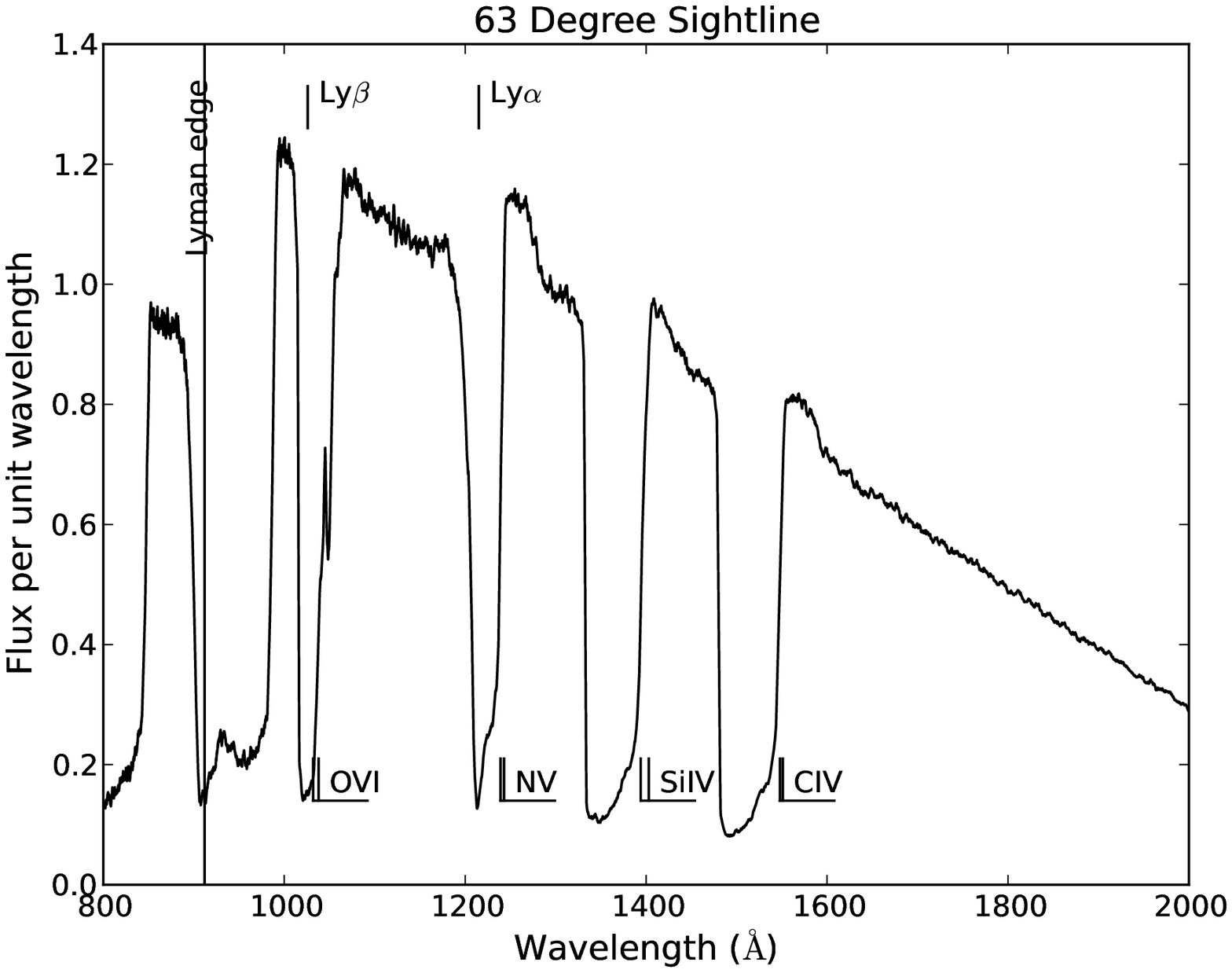}
\plottwo{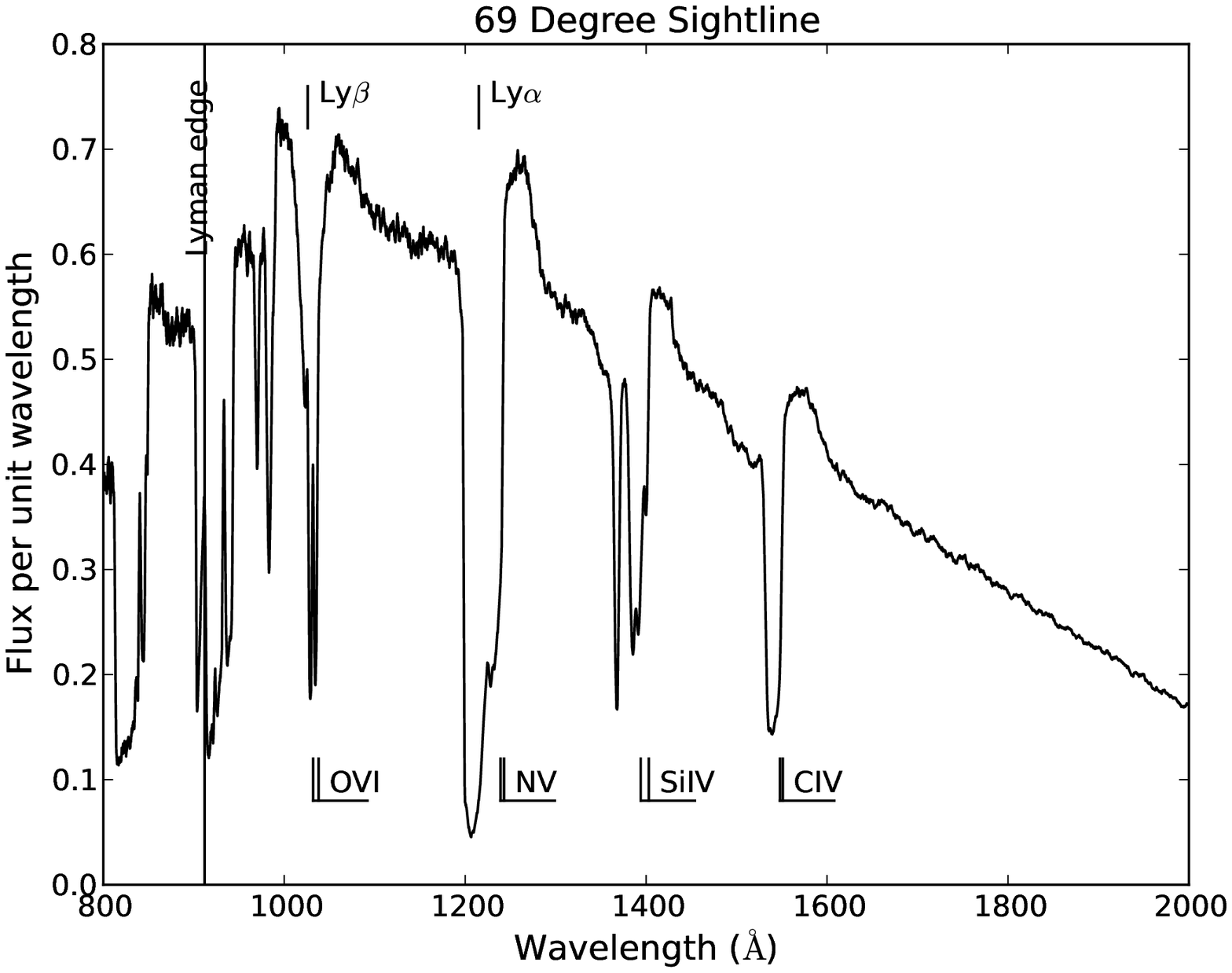}{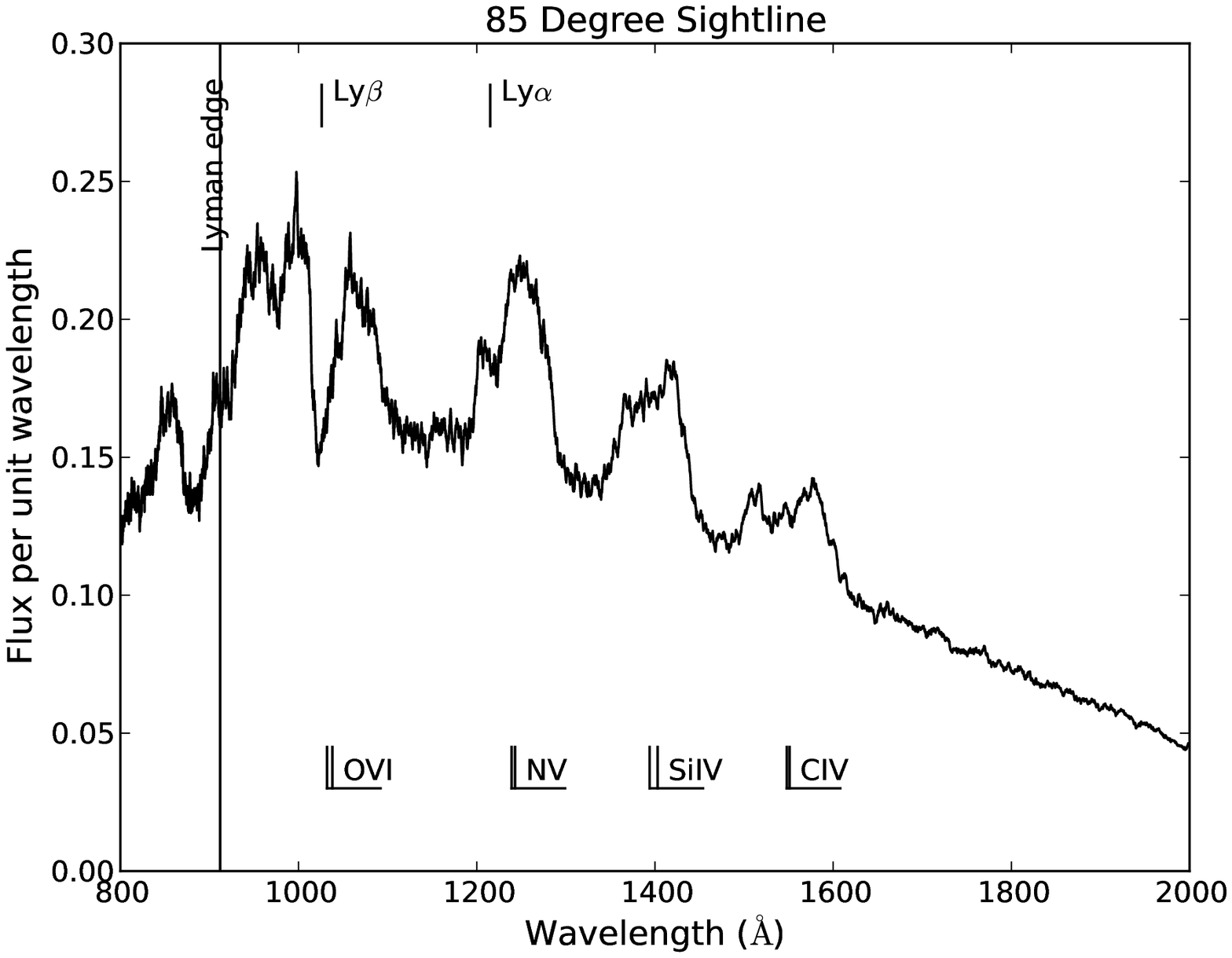}

\caption{Spectra for i=$0^{\circ}$, $63^{\circ}$, $69^{\circ}$ and $85^{\circ}$ calculated for the model shown in figure \ref{wgeom}}
\label{spec}
\end{figure}

In Figure \ref{spec} a we present simulated spectra for observers with inclination angles i=$0^{\circ}$,  $63^{\circ}$,  $69^{\circ}$ and $85^{\circ}$ for this wind model.

The i=$0^{\circ}$ plot shows the calculated spectrum looking down on the disk/wind system, i.e. with the disk seen face on. The wind produces several emission lines, but the bulk of the emission is coming directly from the disk, the central parts of which are unobscured from this direction.
For i=$63^{\circ}$, the observers line of sight towards the centre of the accretion disk lies entirely within the outflow. The wind accelerates along streamlines within this wind cone, and so material at a range of velocities is in the line of sight. C~{\sc iv} is found along much of the top edge of the wind, and so the 1550\AA~ line is visible at a range of velocities, giving rise to the characteristic deep, broad, absorption feature extending blue-ward of the C~{\sc iv} rest wavelength. As shown in Figure \ref{tevel}, O~{\sc vi} is found in regions with lower velocity. The shape of the O~{\sc vi} 1240\AA~ line reflects this: it is much narrower than the carbon line. Features due to other ions commonly used as BALQSO tracers are also seen with similar profiles. The dramatic effect on the line profiles of small changes in viewing angle is evident by comparing the i=$63^{\circ}$ spectrum with that computed for i=$69^{\circ}$.
For i=$85^{\circ}$ the observer is looking under the wind cone, and the C~{\sc iv} line, amongst others, is seen in emission. The line is broadened due to the Keplarian rotation of the wind close to the disk.

\section{Summary and future work}
We have begun a project to use Python, a Monte Carlo radiative transfer code, to simulate the UV spectra of QSO winds. We present initial results for a geometry suggested by \cite{2000ApJ...545...63E}, illuminated by a disk only, and observe that the simulated spectra exhibit features reminiscent of those seen in BALQSOs. 

At present, for the model parameters we have tried, the wind becomes over-ionized whenever a strong central X-ray source is included, and none of the expected UV lines are observed.  We are incorporating a dense shielding region into the geometry to see if this produces BAL-like spectra even with a strong central X-ray source. 

We will produce synthetic spectra for a range of geometries, viewed from different sight-lines to and so inform the discussion of the evolutionary vs orientation interpretation of the BALQSO phenomenon.

\acknowledgements The work presented in this paper is funded by a grant from the Science and Technology Facilities Council (STFC).
\bibliography{higginbottom}
\end{document}